\documentclass{emulateapj}

\newcommand{\etal}{{et al.}}
\newcommand{\zzsun}{[$Z/Z_{\odot}$]}
\newcommand{\tnm}{\tablenotemark}
\newcommand{\tnt}{\tablenotetext}

\begin{document}

\shorttitle{Infrared Spectroscopic Study of 8 Galactic Globular Clusters}
\shortauthors{Stephens \& Frogel}
\journalinfo{The Astronomical Journal, 2004 February}
\submitted{Recieved 2003 May 27; accepted 2003 November 10}

\title{Infrared Spectroscopic Study of 8 Galactic Globular Clusters}

\author{Andrew W. Stephens\altaffilmark{1,3}}
\affil{Pontificia Universidad Cat\'{o}lica de Chile, 
Departamento de Astronom\'{\i}a y Astrof\'{\i}sica, 
Cassilla 306, Santiago 22, Chile; 
and Princeton University Observatory, 
Peyton Hall - Ivy Lane, 
Princeton, NJ 08544-1001; stephens@astro.puc.cl}

\and 

\author{Jay A. Frogel \altaffilmark{2,3,4}}
\affil{NASA Headquarters, 300 E Street SW, Washington, DC  20546}

\altaffiltext{1}{Princeton-Catolica Prize Fellow}

\altaffiltext{2}{Permanent address: The Ohio State University, 
Department of Astronomy, 140 West 18th Avenue, Columbus, OH  43210}

\altaffiltext{3}{Visiting Astronomer, CTIO, which is part of the National
Optical Astronomy Observatory, operated by AURA, Inc., under cooperative
agreement with the NSF.}

\altaffiltext{4}{Visiting Investigator, Department of Terrestrial 
Magnetism, Carnegie Institute of Washington.}

\begin{abstract}

We have obtained medium-resolution infrared $K$ band spectra of 44
giants in seven heavily reddened clusters in the Galactic bulge, as well
as 12 giants in $\omega$~Centauri.  We measure the equivalent widths of
the Na doublet, the Ca triplet, and the CO band head, and then apply the
new technique of Frogel \etal to determine the metallicity of each star.
Averaging these values, we estimate the metallicity for each cluster,
and compare our new [Fe/H] values with previous determinations from the
literature.
Our estimates for each cluster are:
NGC~6256 ($-1.35$), 
NGC~6539 ($-0.79$), 
HP~1 ($-1.30$), 
Liller~1 ($-0.36$), 
Palomar~6 ($-0.52$), 
Terzan~2 ($-0.87$), 
and Terzan~4 ($-1.62$).
We briefly discuss differences between the various [Fe/H] scales on
which it was possible to base our calibration, which is found to be the
largest uncertainty in using this technique to determine metallicities.

\end{abstract}


\section{Introduction} \label{sec:introduction}

The globular clusters of the Milky Way consist of two distinctly
separated groups, each with approximately Gaussian metallicity ([Fe/H])
distributions.  This bimodality was first quantified by
\citet{Freeman1981}, \citet{Zinn1985}, and \citet{Armandroff1988}.  The
mean [Fe/H] values for the two groups are approximately $-0.5$ and
$-1.6$. As detailed by these authors and many others since then, the
bimodality is not confined just to the distribution over [Fe/H].  The
two groups have quite distinct spatial distributions and kinematical
properties as well.  The metal-poor group contains classical, population
II halo objects in terms of their location and space motions, in
addition to their low metallicities.  Clusters in the metal-rich group,
on the other hand, are mostly confined to the same volume of space as
the Galactic bulge.  Of the 33 clusters with [Fe/H] $\ge -0.7$ (data
from \citet{Harris1996} as updated in 2003 February), three-quarters of
them lie within 20\degr\ of the Galactic center, and half lie within
10\degr\ of the center.  Not surprisingly, these clusters are also
heavily reddened; the average $E(B-V)$ of the sample that lies within
10\degr\ of the Center is 1.41, which is equivalent to an average visual
extinction ($A_V$) of 4.4 mag.  The problems associated with optical
studies of these clusters are further compounded by the fact that many
of them lie in very crowded fields and are not themselves of high
surface brightness or concentration.

An obvious question to ask concerning the metal-rich globular clusters
is whether they are in some way physically associated with the bulge,
sharing a common origin and evolution, as has been argued by many
authors \citep[e.g.][]{Minniti1995b, Cote1999}; or is the similarity in
spatial distribution of the clusters and bulge stars merely coincidental
and due simply to the influence of the same gravitational potential.  If
the former is the case, then knowledge of these metal-rich clusters
would be particularly relevant in enhancing our understanding of the
formation and evolution of the bulge itself.  Accurate abundance
determinations for these clusters would be of particular importance.
However, because of the difficulties associated with observing these
clusters in the optical, spectroscopic abundances based on observations
of individual stars are available for only a small fraction of those
that lie closer than 15\degr\ from the Galactic center.  The situation
is somewhat better in terms of deriving physical parameters from optical
color-magnitude diagrams (CMDs); of particular note is the program being
carried out by Ortolani and his collaborators \citep[][and references
therein]{Ortolani1997a}.

In working toward this goal of learning more about these heavily
reddened bulge clusters, we have recently developed a new method for the
determination of [Fe/H] for globular clusters based on the measurement
of absorption features due to Ca, Na, and CO in the $K$ band spectra of
a few of the brightest giants in each cluster \citep[][hereafter
Paper~I]{Frogel2001}. The practical advantages of the near-IR technique
we have developed are several and important. The necessary spectroscopic
observations are at a resolution of only $\approx 1500$ and need only be
obtained for the brightest half-dozen or so giants in a cluster.
Knowledge of cluster reddening or distance is not needed, although if
dereddened colors and absolute $K$ band magnitudes are known, the
accuracy of the resulting [Fe/H] value is further improved.
Furthermore, because of the close coupling between [Fe/H] and the
location and morphology of a cluster's red giant branch (RGB), our
technique is largely insensitive to the choice of stars so long as they
are on the upper 3 or 4 mag ($K$ band) of the RGB.  This advantage was
also demonstrated in our study of giants in the field of the Galactic
bulge \citep{Ramirez2000}.  In the $K$ band the giants near the top of a
cluster's RGB are very bright in contrast to their relative faintness in
$V$, which is due to severe molecular blanketing. And, finally, a
consequence of the previous fact is that the effects of crowding and
confusion with field stars are significantly reduced.

\section{Selection of Clusters, Observations, Data Reduction}\label{sec:selection}

For this paper we chose seven poorly studied globular clusters located
in or near the Galactic bulge, which are believed to have [Fe/H] $\ge
-1.0$ and/or are heavily reddened.  For five of these clusters, our work
is the first detailed spectroscopic study of individual cluster
stars. Table~\ref{tab:clusters} summarizes some of the properties of
these clusters.

\begin{deluxetable*}{lcccccccccc}
\tablewidth{0pt}
\tablecaption{Clusters observed \& some previous metallicity determinations}
\tabletypesize{\footnotesize}
\tablehead{
\colhead{Cluster} 		&
\colhead{{\it l}} 		&
\colhead{{\it b}}		&
\colhead{$E(B-V)$\tnm{a}}	&
\colhead{[Fe/H]\tnm{b}$_{ZW84}$}&
\colhead{[m/H]\tnm{c}$_{W85}$}	&
\colhead{[Fe/H]\tnm{d}$_{AZ88}$}&
\colhead{\zzsun\tnm{e}$_{B98}$}	&
\colhead{[Fe/H]\tnm{a}$_{H03}$}	}
\startdata
NGC 6256  &$-12.2$ &  3.3 & 1.03 & \ldots  & $-1.56$ & \ldots  & $-1.01$ & $-0.70$ \\
NGC 6539  &  20.8  &  6.8 & 0.97 & $-0.66$ & $-1.05$ & \ldots  & \ldots  & $-0.66$ \\
HP 1      & $-2.6$ &  2.1 & 0.74 & \ldots  & $-1.68$ & $-0.56$ & $-1.09$ & $-1.55$ \\
Liller 1  & $-5.2$ &$-0.2$& 3.06 & $-0.21$ & $-0.29$ & $+0.20$ & $+0.08$ & $+0.22$ \\
Palomar 6 &   2.1  &  1.8 & 1.46 & $-0.74$ & $+0.22$ & \ldots  & $-0.09$ & $-1.09$ \\
Terzan 2  & $-3.7$ &  2.3 & 1.57 & $-0.47$ & $-0.54$ & $-0.25$ & $-0.26$ & $-0.40$ \\
Terzan 4  & $-4.0$ &  1.3 & 2.35 & $-0.21$ & $-0.29$ & $-0.94$ & $-0.61$ & $-1.60$ \\
\hline
$\Delta$[Fe/H]\tnm{f}& & & & $0.37\pm 0.62$ & $0.23\pm0.62$ & $0.64\pm0.08$ & $0.50\pm0.28$ & $0.14\pm0.45$ \\
\enddata
\tnt{a}{from the 2003 version of the \citet{Harris1996} catalog}
\tnt{b}{from integrated spectroscopy \citep{Zinn1984} or NIR photometry \citep{Malkan1982}}
\tnt{c}{from dereddened subgiant colors \citep{Webbink1985}}
\tnt{d}{from integrated measurements of the Ca~{\sc ii} triplet \citep{Armandroff1988}}
\tnt{e}{from integrated measurements of the Ca~{\sc ii} triplet \citep{Bica1998}}
\tnt{f}{compared to our final values given in Table~\ref{tab:bestvalues}}
\label{tab:clusters}
\end{deluxetable*}

Most previous estimates of [Fe/H] values for these seven clusters are
from the characteristics of their optical or near-IR CMDs, or from
integrated spectroscopy.  Integrated spectroscopy of globular clusters
that are in the very crowded Galactic bulge region, especially optical
spectroscopy, will almost always tend to overestimate [Fe/H] because of
contamination from metal-rich bulge giants (see the discussion in
Section~\ref{sec:conclusions}).

Our spectroscopic data were obtained on the Blanco 4 m telescope at the
Cerro Tololo Inter-American Observatory (CTIO) with either CTIO's
near-IR spectrometer \citep[IRS; R=1650,][]{DePoy1990} or with the Ohio
State InfraRed Imager/Spectrograph \citep[OSIRIS; R=1380,][]{DePoy1993}.
Spectral coverage was generally between 2.17 and $2.34 \mu $m.  Most of
our observations were made on the same nights as those for the
calibration clusters (Paper~I); details of the observations and data
reduction for these seven clusters are identical to those described in
Paper~I, and so will be only briefly described here.

Each spectrum begins as a set of $\sim 10$ individual spectra taken by
stepping the star along the slit to eliminate the effects of uneven slit
illumination, bad pixels, and fringing.  The slit is several arc minutes
long, which permits simultaneous star and sky measurements.  The array
response signature is first removed by dividing each of the individual
frames by a dome flat field.  The sky frame for each star is formed by
median-combining the individual frames in each set, thus removing all
traces of the star itself, and then subtracting it from each spectral
frame.  Each spectra in a set is then extracted using the IRAF {\sc
apall} routine.  The final spectrum for a star is the average of the
extracted individual spectra.

The wavelength calibration for each star is determined by fitting to
approximately 12 atmospheric OH lines \citep{Oliva1992} visible on a sky
spectrum, where the sky spectrum is extracted from the sky frame using
the program star as an extraction template.  We fit the observed
wavelengths to the laboratory ones with a second order Legendre
polynomial.

To remove telluric absorption features from each final stellar spectrum,
we divide them by a normalized standard star spectrum to correct for
telluric absorption features and then multiply by a 10,000 K blackbody
spectrum to restore the over all shape of the spectrum.  These standard
stars are of early A spectral type or hotter, with no significant
spectral features in the observed wavelength regime, except for Brackett
$\gamma$ at 2.16 $\mu$m.  They are observed as close in time and air
mass as practical to the observations of each cluster star.

The technique we use to measure the equivalent widths (EWs) of the
spectral features -- EW(Na), EW(Ca), and EW(CO) -- is described in
Paper~I.  Table~\ref{tab:wavelengths} gives the continuum and line
regions over which we integrate.  Note that EW(CO) does not include any
contribution from the $^{13}$CO band head.  EW(Ca) and EW(Na) have
continuum regions on either side of the line absorption region and thus
are independent of changes in the spectral slope due to reddening.
There is no viable continuum region to the red of the CO band head, so
instead we extrapolate a linear fit to the continuum blueward of the
feature and integrate the depth of the absorption under this
extrapolated continuum.  Although the slope of the fitted continuum will
change with reddening, the integrated absorption equivalent width under
the continuum should stay constant.

\begin{deluxetable}{lc}
\tablewidth{0pt}
\tablecaption{Equivalent Width Measurement Intervals}
\tabletypesize{\footnotesize}
\tablehead{
\colhead{Feature}               &
\colhead{Wavelengths (\micron)} }
\startdata
Na I feature             &  2.2040 - 2.2107 \\
Na I continuum           &  2.1910 - 2.1966 \\
Na I continuum           &  2.2125 - 2.2170 \\
\hline
Ca I feature             &  2.2577 - 2.2692 \\
Ca I continuum           &  2.2450 - 2.2560 \\
Ca I continuum           &  2.2700 - 2.2720 \\
\hline
$^{12}$CO(2,0) band      &  2.2915 - 2.3025 \\
$^{12}$CO(2,0) continuum &  2.1900 - 2.2010 \\
$^{12}$CO(2,0) continuum &  2.2110 - 2.2220 \\
$^{12}$CO(2,0) continuum &  2.2330 - 2.2600 \\
$^{12}$CO(2,0) continuum &  2.2680 - 2.2800 \\
$^{12}$CO(2,0) continuum &  2.2860 - 2.2910 \\
\enddata
\label{tab:wavelengths}
\end{deluxetable}

Table~\ref{tab:indices} gives the observed spectroscopic indices for
each star, and when available, colors and magnitudes.  The photometry is
from \citet{Frogel1995} for Liller~1, \citet{Kuchinski1995} for
Terzan~2, and Frogel \& Sarajedini (1998, private communication) for
Terzan~4.  Measurement uncertainties for the spectroscopic indices are
typically 0.3, 0.4, and 0.6 \AA\ for Na, Ca, and CO, respectively,
calculated from the errors associated with fitting the continuum.
However, an analysis of duplicate measurements taken over several years
and with different instruments yields slightly higher values (see
Table~3 of Paper~I).

\begin{deluxetable*}{lccccccc}
\tablewidth{0pt}
\renewcommand{\arraystretch}{.6}	
\tablecaption{Near-IR Spectral Indices, Colors, and Luminosities}
\tabletypesize{\footnotesize}
\tablehead{
\colhead{}		&
\colhead{}		&
\colhead{}		&
\colhead{}		&
\multicolumn{3}{c}{\underline{ \hspace{0.9cm} EW (\AA) \hspace{0.9cm}}} &
\colhead{}		 \\
\colhead{Cluster}	&
\colhead{Star}		&
\colhead{$M_{K0}$}	&
\colhead{$(J-K)_0$}	&
\colhead{Na}		&
\colhead{Ca}		&
\colhead{CO}		&
\colhead{[Fe/H]\tnm{c}}	}
\startdata
NGC 6256 &  51\tnm{a} & \ldots & \ldots &  0.60 &  0.71 &  6.52 & -1.37 \\ 
NGC 6256 &  93 & \ldots & \ldots &  0.67 &  0.85 &  5.51 & -1.37 \\ 
NGC 6256 &  95 & \ldots & \ldots &  0.79 &  0.37 &  3.37 & -1.41 \\ 
NGC 6256 & 162 & \ldots & \ldots &  1.01 &  0.40 &  4.20 & -1.31 \\ 
NGC 6256 &   B & \ldots & \ldots &  0.71 &  0.46 &  8.05 & -1.29 \\ 
\hline
NGC 6539 &   1 & \ldots & \ldots &  1.99 &  2.84 & 16.83 & -0.67 \\ 
NGC 6539 &   2 & \ldots & \ldots &  1.88 &  2.35 & 15.27 & -0.76 \\ 
NGC 6539 &   3 & \ldots & \ldots &  1.85 &  2.71 & 16.45 & -0.72 \\ 
NGC 6539 &   4 & \ldots & \ldots &  1.31 &  1.55 & 12.34 & -0.99 \\ 
\hline
HP 1 &  46 & \ldots & \ldots &  0.63 &  0.93 &  5.73 & -1.38 \\ 
HP 1 &  57 & \ldots & \ldots &  4.80 &  4.30 & 22.69 & -0.25 \\ 
HP 1 & 191 & \ldots & \ldots &  4.63 &  4.01 & 21.35 & -0.29 \\ 
HP 1 & 435 & \ldots & \ldots &  0.62 &  1.21 & 11.92 & -1.22 \\ 
HP 1 &   A & \ldots & \ldots &  4.03 &  2.98 & 21.14 & -0.42 \\ 
HP 1 &   B & \ldots & \ldots &  2.83 &  3.22 & 18.50 & -0.48 \\ 
\hline
Liller 1 &   6 & -6.50 & 1.17 &  3.57 &  4.63 & 21.43 & -0.16 \\ 
Liller 1 &   7 & -6.45 & 1.13 &  2.35 &  5.34 & 14.53 & -0.59 \\ 
Liller 1 & 157 & -6.80 & 0.93 &  4.22 &  5.62 & 21.23 & -0.23 \\ 
Liller 1 & 158 & -6.62 & 1.13 &  3.15 &  4.32 & 20.93 & -0.32 \\ 
Liller 1 & 160 & -6.50 & 1.27 &  2.22 &  2.04 & 12.28 & -0.66 \\ 
Liller 1 & 162 & -6.05 & 1.16 &  2.46 &  3.27 & 18.00 & -0.45 \\ 
Liller 1 & 166 & -5.47 & 0.95 &  3.07 &  3.16 & 11.58 & -0.56 \\ 
Liller 1 & 299 & -6.73 & 1.38 &  3.93 &  5.34 & 21.88 &  0.07 \\ 
\hline
Palomar 6 & 1\tnm{b} & \ldots & \ldots &  3.81 &  4.71 & 21.42 & -0.16 \\ 
Palomar 6 &   2 & \ldots & \ldots &  2.65 &  2.94 & 18.01 & -0.54 \\ 
Palomar 6 & 3\tnm{b} & \ldots & \ldots &  0.97 & -1.14 & 12.64 & -1.03 \\ 
Palomar 6 &   4 & \ldots & \ldots &  1.11 & -1.94 & 17.02 & -0.85 \\ 
Palomar 6 &   5 & \ldots & \ldots &  4.64 &  3.04 & 22.40 & -0.43 \\ 
Palomar 6 &  11 & \ldots & \ldots &  4.55 &  4.75 & 18.54 & -0.16 \\ 
Palomar 6 &  14 & \ldots & \ldots &  2.78 &  1.77 & 17.33 & -0.63 \\ 
\hline
Terzan 2 &   1 & -5.82 & 0.82 &  1.37 &  2.18 & 12.45 & -1.07 \\ 
Terzan 2 &   2 & -5.78 & 0.95 &  1.02 &  2.39 & 17.15 & -0.83 \\ 
Terzan 2 &   3 & -5.73 & 0.99 &  4.86 &  4.03 & 22.39 &  0.27 \\ 
Terzan 2 &   4 & -5.72 & 0.85 &  1.93 &  2.36 & 18.16 & -0.74 \\ 
Terzan 2 &   5 & -5.46 & 0.97 &  0.83 &  2.85 & 16.67 & -0.79 \\ 
Terzan 2 &   7 & -5.36 & 0.88 &  1.32 &  3.23 & 16.58 & -0.76 \\ 
Terzan 2 &   8 & -5.33 & 0.87 &  0.82 &  1.83 & 11.67 & -1.04 \\ 
\hline
Terzan 4 & 185 & -6.44 & 0.74 &  0.97 &  0.91 &  6.28 & -1.49 \\ 
Terzan 4 & 333 & -6.25 & 0.84 &  1.16 & -0.15 &  7.67 & -1.33 \\ 
Terzan 4 & 399 & -6.69 & 0.73 & -0.07 &  1.96 & -0.78 & -1.90 \\ 
Terzan 4 & 424 & -6.11 & 0.69 & -0.01 &  2.02 &  1.45 & -1.77 \\ 
Terzan 4 & 453 & -6.61 & 0.85 &  0.01 &  0.85 &  8.70 & -1.57 \\ 
Terzan 4 & 461 & -5.92 & 0.70 & -0.33 &  0.31 &  3.18 & -1.80 \\ 
Terzan 4 & 474 & -6.22 & 0.78 &  0.35 &  0.89 &  9.20 & -1.48 \\ 
\enddata
\tnt{a}{The average of measurements taken on two nights.}
\tnt{b}{Non-members based on the radial velocities of \citet{Lee2002}.}
\tnt{c}{Based on the QQ3 solution for NGC~6256, NGC~6539, HP~1, 
and Palomar~6, while Liller~1 and Terzan~4 are based on the LLA9
solution, and Terzan~2 is based on the QQA9 solution (see Table
\ref{tab:solutions}).}
\renewcommand{\arraystretch}{1}
\label{tab:indices}
\end{deluxetable*}

\begin{figure}[htb]
\epsscale{1.1}
\plotone{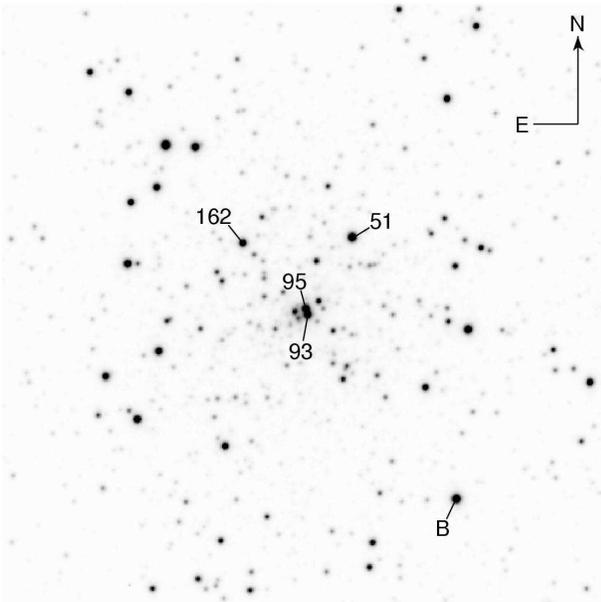} 
\figcaption{
The $2' \times 2'$ $I$ band finder chart for the stars observed in
NGC~6256.
\label{fig:ngc6256}}
\end{figure}

\begin{figure}[htb]
\epsscale{1.1}
\plotone{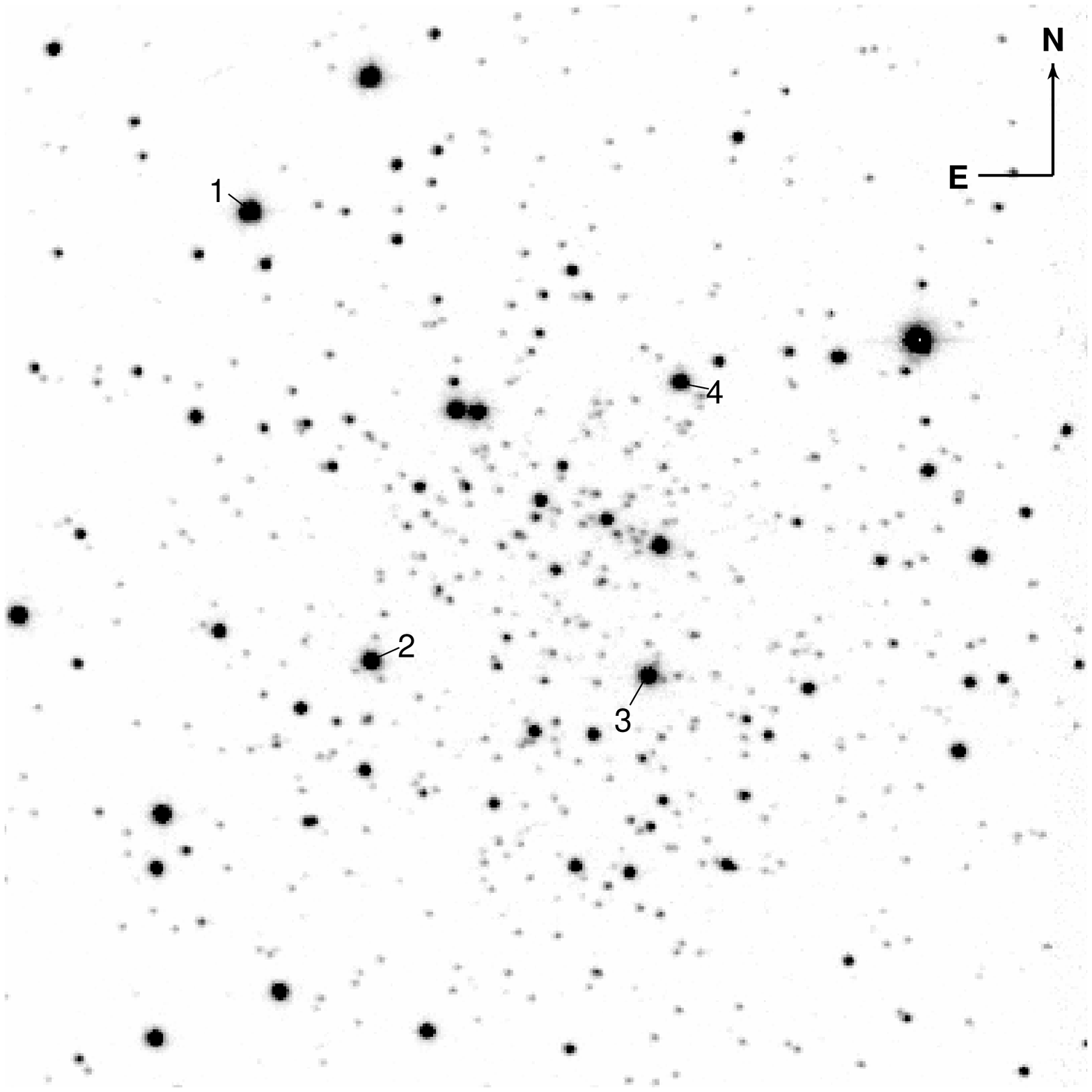} 
\figcaption{
The $2' \times 2'$ $K$ band finder chart for the stars observed in
NGC~6539.
\label{fig:ngc6539}}
\end{figure}

\begin{figure}[htb]
\epsscale{1.1}
\plotone{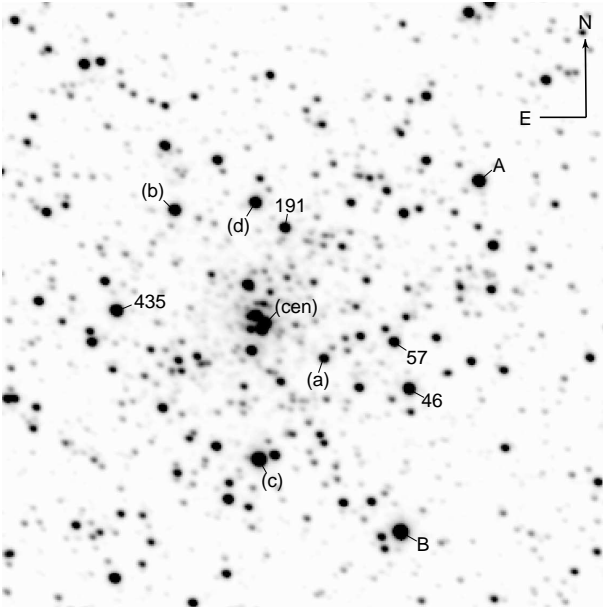} 
\figcaption{
The $1.5' \times 1.5'$ $I$ band finder chart for the stars observed in
HP~1.  The letters in parentheses mark stars observed spectroscopically
by \citet{Minniti1995a}(a,b,c,d,cen).
\label{fig:hp1}}
\end{figure}

\begin{figure}[htb]
\epsscale{1.1}
\plotone{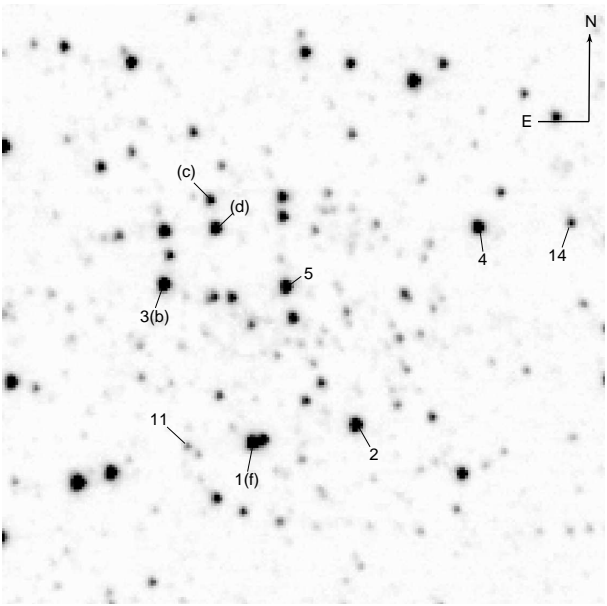} 
\figcaption{
The $2' \times 2'$ $K$ band finder chart for the stars observed in
Palomar~6.  The letters in parentheses mark stars observed
spectroscopically by \citet{Lee2002} (a-g), and by
\citet{Minniti1995a}(b,c,d,f).
\label{fig:pal6}}
\end{figure}

\begin{figure}[htb]
\epsscale{1.1}
\plotone{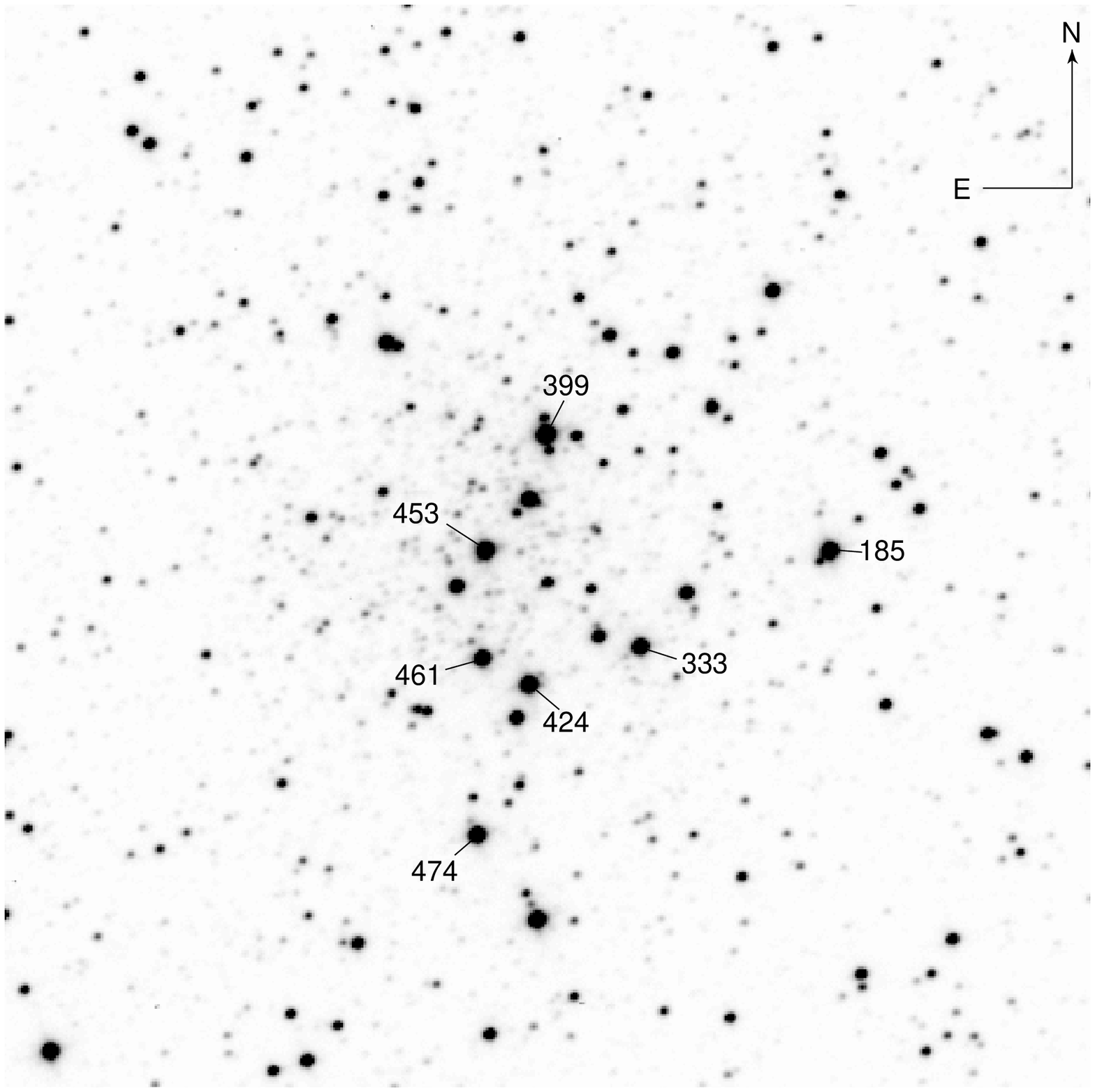} 
\figcaption{
The $1' \times 1'$ $I$ band finder chart for the stars observed in
Terzan~4.
\label{fig:ter4}}
\end{figure}

Column~2 of Table~\ref{tab:indices} lists the stars observed in each
cluster.  The identifications for Liller~1 are from \citet{Frogel1995},
and the identifications for Terzan~2 are from \citet{Kuchinski1995}.
The numbering schemes for the remaining clusters are completely
arbitrary. We therefore provide finding charts for NGC~6256 in
Figure~\ref{fig:ngc6256}, NGC~6539 in Figure~\ref{fig:ngc6539}, HP~1 in
Figure~\ref{fig:hp1}, Palomar~6 in Figure~\ref{fig:pal6}, and Terzan~4
in Figure~\ref{fig:ter4}.
The last column of Table~\ref{tab:indices} lists the metallicity derived
for each star as discussed in the next section.

For clusters without calibrated near-IR photometry (either our own or
that of others), we constructed instrumental color-magnitude diagrams
from new $JK$ images taken under non-photometric conditions, and chose
the brightest stars near the cluster centers with colors and magnitudes
consistent with membership.  For clusters that physically lie within
particularly crowded bulge fields, this may still not preclude the
inclusion of non-members, since bulge giants will suffer from the same
degree of reddening and will have colors and magnitudes similar to those
of giants in the clusters, especially the metal rich clusters.
Table~\ref{tab:meanvalues} gives the cluster mean values for the indices
and photometric parameters, along with the star-to-star intra cluster
dispersions. The two entries for HP~1 will be explained later.

\begin{deluxetable*}{lccccccccccc}	
\tablewidth{0pt}
\tablecaption{Cluster Mean Values and Dispersions for Observed Parameters}
\tabletypesize{\footnotesize}
\tablehead{
\colhead{}			&
\colhead{}			&
\multicolumn{2}{c}{\underline{ \hspace{0.2cm} EW(Na) \hspace{0.2cm} }} 	&
\multicolumn{2}{c}{\underline{ \hspace{0.2cm} EW(Ca) \hspace{0.2cm} }} 	&
\multicolumn{2}{c}{\underline{ \hspace{0.2cm} EW(CO) \hspace{0.2cm} }} 	&
\multicolumn{2}{c}{\underline{ \hspace{0.2cm} $M_{K0}$ \hspace{0.2cm} }} &
\multicolumn{2}{c}{\underline{ \hspace{0.2cm} $(J-K)_0$ \hspace{0.2cm} }} \\
\colhead{Cluster}		&
\colhead{$N_{stars}$}		&
\colhead{mean}			&
\colhead{$\sigma$}		&
\colhead{mean}			&
\colhead{$\sigma$}		&
\colhead{mean}			&
\colhead{$\sigma$}		&
\colhead{mean}			&
\colhead{$\sigma$}		&
\colhead{mean}			&
\colhead{$\sigma$}		}
\startdata
NGC 6256    & 5 & 0.76 & 0.16 & 0.56 & 0.21 &  5.53 & 1.86 & \ldots & \ldots &\ldots & \ldots \\
NGC 6539    & 4 & 1.76 & 0.30 & 2.36 & 0.58 & 15.22 & 2.03 & \ldots & \ldots &\ldots & \ldots \\
HP 1 weak   & 2 & 0.63 & 0.01 & 1.07 & 0.20 &  8.83 & 4.38 & \ldots & \ldots &\ldots & \ldots \\
HP 1 strong & 4 & 4.07 & 0.89 & 3.63 & 0.63 & 20.92 & 1.75 & \ldots & \ldots &\ldots & \ldots \\
Liller 1    & 8 & 3.12 & 0.75 & 4.21 & 1.28 & 17.73 & 4.33 & -6.39  & 0.44   &  1.14 & 0.15   \\
Palomar 6   & 5 & 3.15 & 1.48 & 2.11 & 2.50 & 18.66 & 2.17 & \ldots & \ldots &\ldots & \ldots \\
Terzan 2    & 7 & 1.74 & 1.43 & 2.70 & 0.74 & 16.44 & 3.60 & -5.60  &  0.21  &  0.90 & 0.07   \\
Terzan 4    & 7 & 0.30 & 0.56 & 0.97 & 0.80 &  5.10 & 3.86 & -6.32  &  0.27  &  0.76 & 0.06   \\
\enddata
\tablecomments{
Pal~6 Excludes the 2 known non-members.
}
\label{tab:meanvalues}
\end{deluxetable*}

\begin{figure}[htb]
\epsscale{1.2}
\plotone{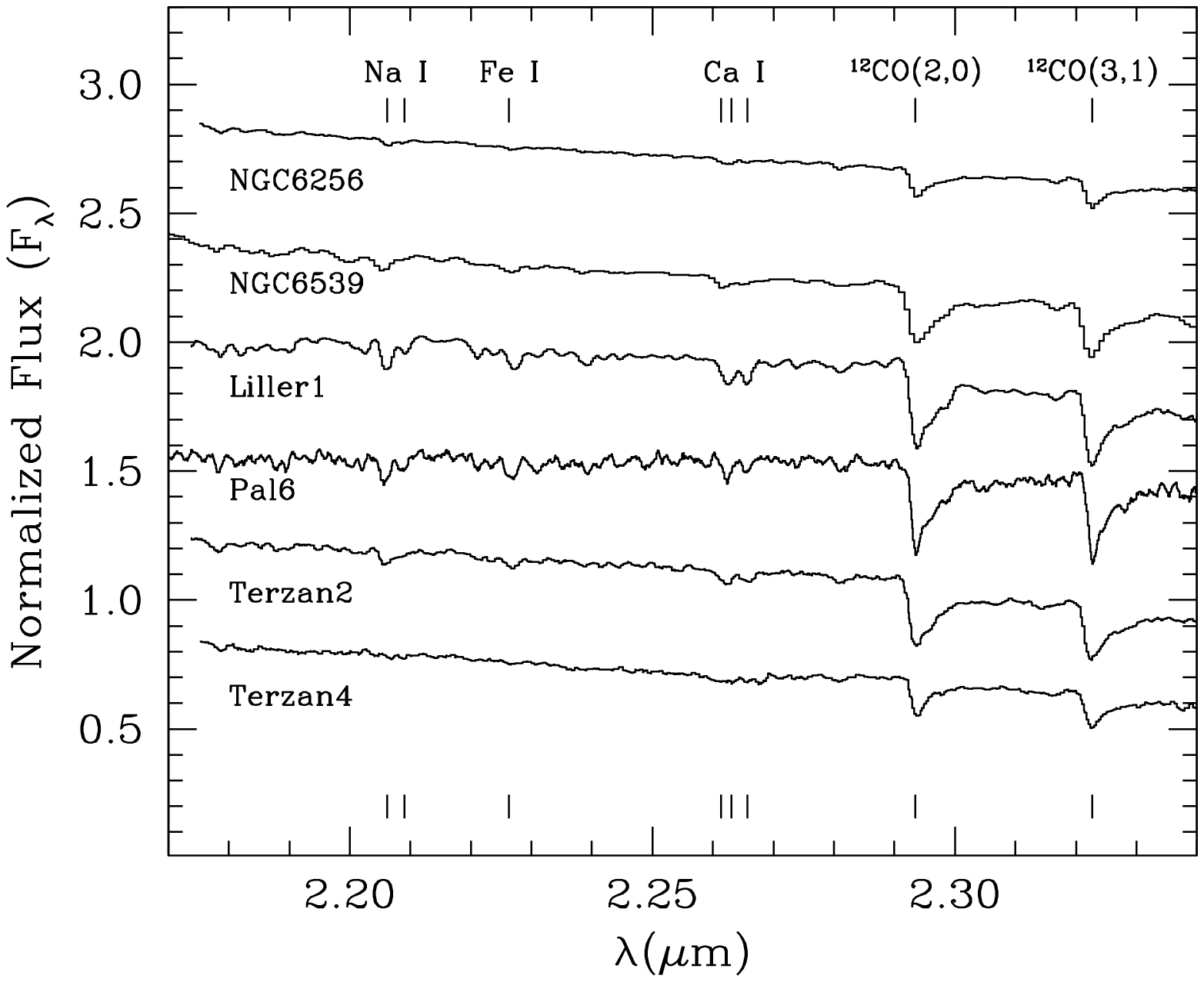} 
\figcaption{
Average of the individual stellar spectra for each cluster observed,
except for HP~1 (see Fig. \ref{fig:hp1spec}).  The positions of the
Na~{\sc i} doublet, a strong Fe~{\sc i} line, the Ca~{\sc i} triplet,
and the first two band heads of CO are indicated.  Only the
$^{12}$CO(2,0) band is used for determining EW(CO).  Note the relative
absence of spectral features in NGC~6256 and Terzan~4.  The high S/N of
the spectra is also evident.
\label{fig:spectra}}
\end{figure}

Figure~\ref{fig:spectra} illustrates, for each cluster except HP~1, the
average of the individual stellar spectra.  Here we have indicated the
three features measured, the Na~{\sc i} doublet, the Ca~{\sc i} triplet,
and the first band head of CO.  For reference, we have also marked a
strong Fe~{\sc i} line and the second band head of CO.

\begin{figure}[htb]
\epsscale{1.2}
\plotone{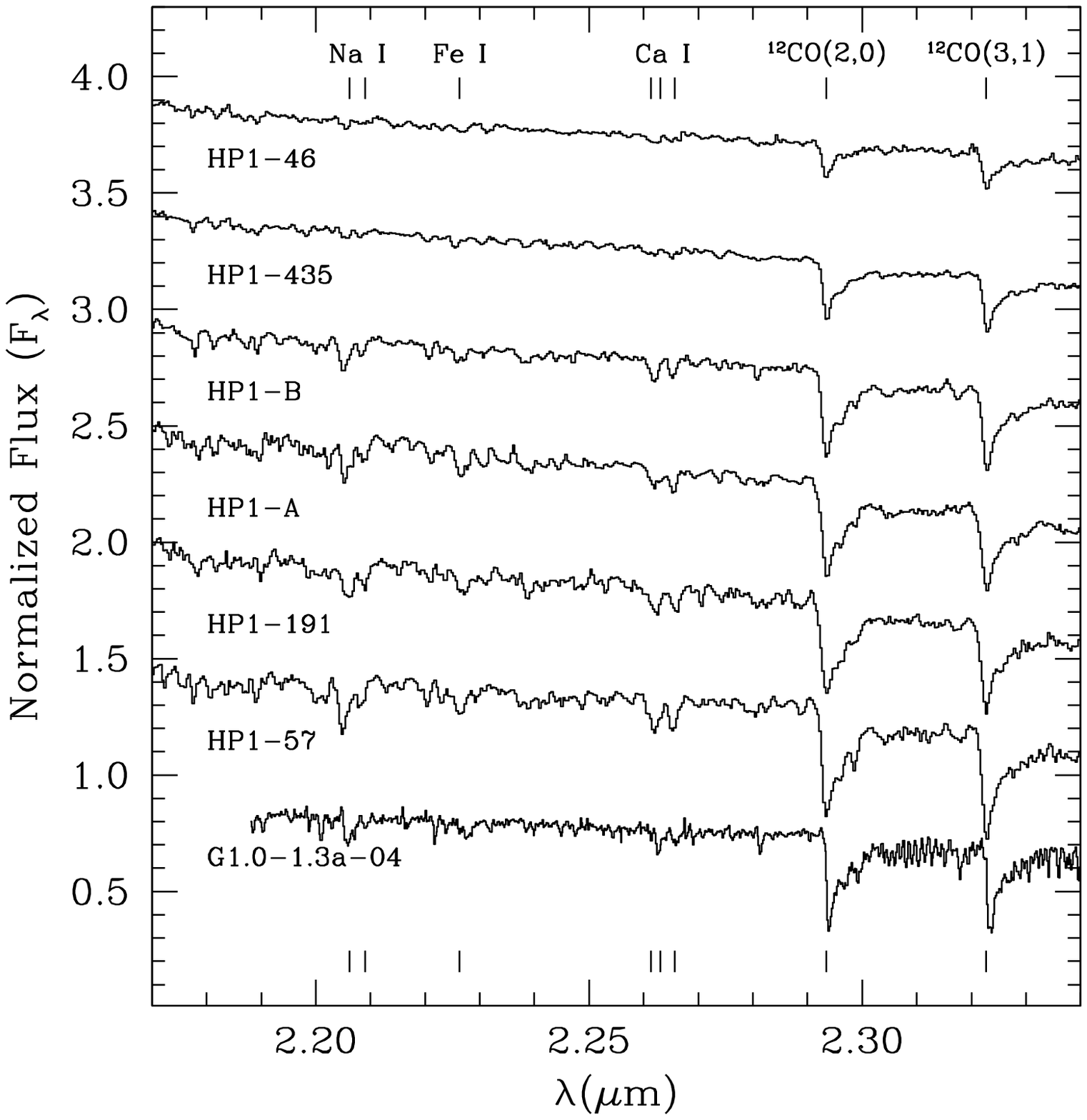} 
\figcaption{
$K$ band spectra of the six stars observed in HP~1, and the spectrum
of an inner galactic bulge star with similar colors and luminosity for
comparison \citep{Ramirez2000}.  The first two stars, 46 and 435,
clearly have much weaker features than the other four stars in HP~1.
These first two stars are therefore referred to as the ``weak'' group,
while the latter four are referred to as the ``strong'' group.
\label{fig:hp1spec}}
\end{figure}

Figure~\ref{fig:hp1spec} shows the individual stellar spectra for all
stars observed in HP~1.  Note that the first two of the HP~1 stars have
significantly weaker absorption features than the other four stars.  For
comparison, we also show the spectrum of a low-latitude Galactic bulge
star with similar colors and indices as the ``strong'' HP~1 stars from
\citet{Ramirez2000}.

\section{Derivation of Cluster Metallicities} \label{sec:metallicities}

Paper~I investigated techniques to use spectroscopic features measured
in medium-resolution infrared spectra to measure globular cluster
metallicities.  The three strongest features in the $K$ band spectrum
are the Na~{\sc i} doublet, the Ca~{\sc i} triplet, and the first band
head of CO.  

It is well known that each of these features has its own problems as a
metallicity indicator.  [Na/Fe] can show significant star-to-star
variations within a single cluster; \citet{Sneden2000} have measured
$\sigma$[Na/Fe]$\simeq 0.25$ dex in M15 and M92.  Calcium is an
$\alpha$-element, and thus suffers from $\alpha$-enhancement, which
varies from cluster to cluster and systematically with overall
abundance \citep{Carney1996}.  The first-overtone CO band heads in the
$K$ band saturate at high metallicities, making them less sensitive to
carbon abundance and more dependent on the microturbulent velocity.

While each of these features has its own shortcomings as a metallicity
indicator, taken together and averaged over several stars in a cluster,
we have demonstrated that they nonetheless provide a precise metallicity
determination (Paper~I).  Including stellar luminosities and colors,
which help constrain the stellar temperature and gravity, improves the
technique even further.

Taking advantage of these facts, in Paper~I we derived four equations
with which one can calculate [Fe/H] for globular cluster giants.  These
equations use either linear or quadratic combinations of the measured
equivalent widths of Na, Ca, and CO, and, optionally, the infrared color
and luminosity (Table~10 of Paper~I).  These equations are applied to
each observed star in a cluster to predict [Fe/H] for that star.  The
unweighted average of these values is then the [Fe/H] assigned to the
cluster.  Solutions LL3 and QQ3, linear and quadratic, respectively, use
only the three spectroscopic indices; therefore, they are independent of
reddening and distance to the clusters.  Solutions LLA9 and QQA9 also
incorporate the $(J-K)_0$ colors and $M_K$ magnitudes.  Thus, while
[Fe/H] estimates from LLA9 and QQA9 use more information and can be more
precise, they are also affected by uncertainties in the reddening and
distance estimates.  Table~\ref{tab:solutions} lists the coefficients
for the four equations; it is an abbreviated version of Table~10 from
Paper~I.  The last line of Table~\ref{tab:solutions} is our estimate of
the typical uncertainty of the metallicity derived from each solution.
It is based on the scatter of the individual stellar determinations in
each of the calibrating clusters with six to nine stars observed per
cluster.  It does not include any uncertainties introduced by the
observations of clusters to which the calibration is being applied,
$e.g.$ errors in calculating the distance modulus or reddening for the
solutions that employ colors and absolute magnitudes.  Nor does it
include the absolute uncertainty of the metallicity scale, which is
discussed in section~\ref{sec:fehscale}.

\begin{deluxetable*}{lcccccccc}
\tablewidth{0pt}
\tablecaption{The Linear and Quadratic Solutions for [Fe/H]}
\tabletypesize{\footnotesize}
\tablehead{
\colhead{}			&
\multicolumn{2}{c}{\underline{ \hspace{0.5cm} LL3 \hspace{0.5cm} }} &
\multicolumn{2}{c}{\underline{ \hspace{0.5cm} LLA9 \hspace{0.5cm} }} &
\multicolumn{2}{c}{\underline{ \hspace{0.5cm} QQ3 \hspace{0.5cm} }} &
\multicolumn{2}{c}{\underline{ \hspace{0.5cm} QQA9 \hspace{0.5cm} }} \\
\colhead{variable}		&
\colhead{coef}			&
\colhead{$\sigma$}		&
\colhead{coef}			&
\colhead{$\sigma$}		&
\colhead{coef}			&
\colhead{$\sigma$}		&
\colhead{coef}			&
\colhead{$\sigma$}		}
\startdata
const       & -1.663 &  0.057 & -1.451 &  0.18  & -1.811 &  0.074 &  1.097 & 1.16  \\
EW(Na)      &  0.182 &  0.029 &  0.202 &  0.036 &  0.389 &  0.065 &  0.130 & 0.091 \\
EW(Na)$^2$  & \ldots & \ldots & \ldots & \ldots & -0.047 &  0.013 &  0.016 & 0.020 \\
EW(Ca)      &  0.057 &  0.025 &  0.025 &  0.026 & -0.030 &  0.051 &  0.058 & 0.056 \\
EW(Ca)$^2$  & \ldots & \ldots & \ldots & \ldots &  0.024 &  0.012 & -0.004 & 0.014 \\
EW(CO)      &  0.027 &  0.005 &  0.026 &  0.005 &  0.043 &  0.013 &  0.028 & 0.016 \\
EW(CO)$^2$  & \ldots & \ldots & \ldots & \ldots & -0.001 &  0.000 &  0.000 & 0.001 \\
$(J-K)_0$   & \ldots & \ldots &  0.749 &  0.23  & \ldots & \ldots &  5.240 & 1.84  \\
$(J-K)_0^2$ & \ldots & \ldots & \ldots & \ldots & \ldots & \ldots & -2.313 & 0.93  \\
$M_{K}$     & \ldots & \ldots &  0.151 &  0.033 & \ldots & \ldots &  1.812 & 0.40  \\
$M_{K}^2$   & \ldots & \ldots & \ldots & \ldots & \ldots & \ldots &  0.147 & 0.034 \\
\hline
$\sigma$ (est)& 0.11 &        &   0.07 &        &  0.09  &        &  0.06  &       \\
\enddata
\label{tab:solutions}
\end{deluxetable*}

Table~\ref{tab:abundances} lists the average abundances calculated for
each cluster from the four solutions in Table~\ref{tab:solutions}.  The
second column indicates the number of stars observed in each cluster.
The sigma is the star-to-star scatter for each cluster.  While there are
no significant differences in the means between the [Fe/H] values from
any of the solutions, based on the estimated error and the observed star
to star scatter for each solution, we choose a ``best'' metallicity for
each cluster and give the results in Table~\ref{tab:bestvalues}.  These
are computed on a cluster-by-cluster basis, as described in
section~\ref{sec:notes}.  The metallicity of each star, as used in the
calculation of the ``best'' value for each cluster, is also listed in
the last column of Table~\ref{tab:indices}.

\begin{deluxetable*}{lccccccccc}
\tablewidth{0pt}
\tablecaption{Cluster Abundances}
\tabletypesize{\footnotesize}
\tablehead{
\colhead{}			&
\colhead{}			&
\multicolumn{2}{c}{\underline{ \hspace{0.3cm} LL3 \hspace{0.3cm} }} 	&
\multicolumn{2}{c}{\underline{ \hspace{0.3cm} LLA9 \hspace{0.3cm} }}	&
\multicolumn{2}{c}{\underline{ \hspace{0.3cm} QQ3 \hspace{0.3cm} }} 	&
\multicolumn{2}{c}{\underline{ \hspace{0.3cm} QQA9 \hspace{0.3cm} }}	\\
\colhead{Cluster}		&
\colhead{$N_{stars}$}		&
\colhead{[Fe/H]}		&
\colhead{$\sigma$}		&
\colhead{[Fe/H]}		&
\colhead{$\sigma$}		&
\colhead{[Fe/H]}		&
\colhead{$\sigma$}		&
\colhead{[Fe/H]}		&
\colhead{$\sigma$}		}
\startdata 
NGC 6256    & 5 & -1.34 & 0.04 & \ldots & \ldots & -1.35 & 0.05 & \ldots & \ldots \\
NGC 6539    & 4 & -0.80 & 0.14 & \ldots & \ldots & -0.79 & 0.14 & \ldots & \ldots \\
HP 1 weak   & 2 & -1.25 & 0.13 & \ldots & \ldots & -1.30 & 0.11 & \ldots & \ldots \\
HP 1 strong & 4 & -0.15 & 0.24 & \ldots & \ldots & -0.36 & 0.11 & \ldots & \ldots \\
Liller 1    & 8 & -0.38 & 0.29 & -0.36  &  0.25  & -0.31 & 0.28 & -0.30  &  0.29  \\
Palomar 6   & 5 & -0.47 & 0.44 & \ldots & \ldots & -0.52 & 0.25 & \ldots & \ldots \\
Terzan 2    & 7 & -0.75 & 0.38 & -0.77  &  0.40  & -0.83 & 0.27 & -0.71  &  0.45  \\
Terzan 4    & 7 & -1.42 & 0.15 & -1.62  &  0.21  & -1.53 & 0.28 & -1.60  &  0.23  \\
\enddata
\tablecomments{
Pal~6 Excludes the 2 known non-members.
}
\label{tab:abundances}
\end{deluxetable*}

\subsection{How many Stars are Enough?} \label{sec:howmany}

One may ask how many stars are required to obtain a reliable measure of
the metallicity of a cluster.  In the clusters presented here we have
tried to obtain spectra of more than 5 stars in each; however, because
of time or weather constraints, this may not always have been possible;
for example, in NGC~6539 we only have spectra of four stars.  In such
cases, where only a few stars have been measured, it becomes desirable to
know with what accuracy and precision one can determine the cluster
metallicity.  Is it possible, for example, to get away with only
measuring one or two or three stars with our technique?

One important consideration is the potential contamination from
superposed field stars.  With a large sample of spectra it is possible
to effectively reject a few non-members if their metallicities are
sufficiently different from that of the cluster.  However, with a
reduced sample size it becomes impossible to distinguish between members
and nonmembers.

A way to test the robustness of our metallicity determinations against
small sample sizes is to run Monte Carlo simulations on some of the
clusters that have larger samples, and therefore relatively
well-determined metallicities.  To simulate the results of only
measuring $N$ stars in a cluster, we randomly select stars from the
observed sample and average their metallicities to determine that of the
cluster.  By repeating this process $100,000$ times, we accumulate the
distribution of errors, and can therefore estimate the accuracy and
precision with which one can determine the cluster metallicity using
only a sample of $N$ stars.

Two illustrative examples are the spectra of the clusters Terzan~2 and
Terzan~4.  In Terzan~4 we measured seven stars that have a reasonably
narrow distribution of metallicities; therefore, picking even a few
stars gives a reasonable result for the cluster.  On average, only
measuring one star will give the correct answer, but the distribution of
results has a sigma of 0.19 dex.  Averaging the results from two stars,
the width of the distribution is reduced to 0.12 dex, and with three
stars it drops to 0.09 dex.


Terzan~2 also has a relatively narrow distribution of metallicities,
except that, of the seven measured stars, one appears to be a field star
with a much higher metallicity.  In our analysis in Sec~\ref{sec:ter2}
we recognize this fact, and disregard the outlier when determining our
``best'' value for the cluster.  However, with a smaller sample, it
would be much more difficult to recognize contaminants.  This is shown
in the simulations, where on average, the deviation from the ``best''
value is 0.16 dex.  Only measuring one star yields a distribution with a
sigma of 0.42 dex; the mean of two stars give a sigma of 0.27 dex, three
a width of 0.20 dex, etc.


Thus, the appropriate number of spectra for an accurate measure of the
cluster metallicity (where accurate means to within the accuracy of the
technique) appears to be most dependant on the severity of the
non-member contamination.  If one expects no contaminants, e.g. an
isolated cluster, three stars are sufficient to reduce the stochastic
noise inherent when using spectra of the signal-to-noise ratio (S/N)
presented here (consistent with what was found in Paper I).  However, in
more polluted regions, such as the Galactic disk or bulge, more stars
are required to be able to effectively reject non-members.  Furthermore,
the number of stars required increases as the difference between the
cluster metallicity and the contaminating population decreases.

\subsection{The Metallicity Scale} \label{sec:fehscale}

The calibration of the current technique used the metallicities listed
in the 1999 version of the Harris catalog \citep{Harris1996} of Galactic
globular clusters, which is a continually changing average of parameters
taken from literature.  Thus, while Harris started out on the
\citet{Zinn1984} scale, the catalog has been evolving as more and more
values from high-resolution spectroscopy are incorporated.  However
there exist significant discrepancies observed between metallicities
obtained with various techniques, and on the absolute calibration of the
metallicity scale itself.

The \citet[][hereafter ZW84]{Zinn1984} metallicity scale was one of the
first such metallicity scales.  Calibrated primarily from the
spectroscopy of Cohen \citep[see][and references therein]{Cohen1983}, it
used photometric indices measured from cluster integrated light, and was
applied to a total of 121 clusters.  This scale has been updated by
\citet{Rutledge1997} using low-resolution spectroscopic measurements of
the red Ca~{\sc II} triplet in individual giants.

The \citet[][hereafter CG97]{Carretta1997} [Fe/H] scale is based on a
uniform analysis of high-resolution spectra of more than 160 bright red
giants in 24 globular clusters.  Figure~5 of CG97 shows a comparison
between their system and that of ZW84: a nonlinear relation where the
ZW84 values are $\sim 0.1$ dex too high at the metal-rich end
([Fe/H]$>-1$), and $\sim 0.2$ dex too low for clusters of intermediate
metallicity ($-2<$[Fe/H]$<-1$).

The recent work by \citet[][hereafter KI03]{Kraft2003}, using
measurements of Fe~{\sc II} in high-resolution spectra of giants in 16
clusters, confirms that the ZW84 [Fe/H] scale is nonlinear with respect
to the true [Fe/H] scale.  Their Figure~3 shows the clearly nonlinear
relation between theirs and the ZW84 scale, regardless of the model
atmosphere they choose.  However, Figure~2 of KI03 illustrates the
comparison of their scale with that of CG97, showing that the CG97
values are too high, with an offset of $\sim 0.2$ dex at the lowest
metallicities ([Fe/H]$=-2.4$), and agreement at the highest
([Fe/H]$=-0.7$).

\begin{figure}[htb]
\epsscale{1.2}
\plotone{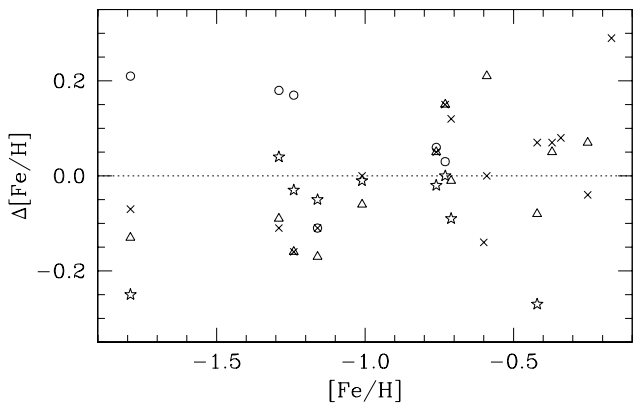} 
\figcaption{
Comparison of our calibration globular cluster metallicities as derived
on different scales, as a function of the metallicity used in the
calibration of our infrared spectroscopic technique.  The difference is
in the sense that if a cluster has a higher metallicity compared with
what we used in our calibration, it will show a positive $\Delta$
[Fe/H].  The crosses represent the clusters listed in \citet{Zinn1984},
the triangles are clusters from \citet{Rutledge1997}, the circles are
from \citet{Carretta1997}, and the stars are from \citet{Kraft2003}.
\label{fig:calib}}
\end{figure}

Since our calibration is not strictly based on any one of these three
different metallicity scales, we have performed a comparison of the
metallicities of the 15 clusters used to calibrate our technique.  The
results of this comparison are shown in Figure~\ref{fig:calib}.  This
shows that at low metallicities, CG97 gives values that tend to be higher
than our calibration, and KI03 gives values that tend to be lower than
our calibration.  At higher metallicities, we are lower than ZW84, but
not as low as KI03.

Table~\ref{tab:calib} summarizes the data in Figure~\ref{fig:calib}; it
includes the metallicity scale, the number of calibrating clusters
measured on each system, the mean difference in metallicity, and the
dispersion of the difference in columns (1) through (4) respectively. In
the mean, our values lie closest to those of ZW84 and Rut97, while they
are intermediate between those of CG97 and KI03.

\begin{deluxetable}{lccc}
\tablewidth{0pt}
\tablecaption{Comparison of Calibrating Clusters}
\tabletypesize{\footnotesize}
\tablehead{
\colhead{Scale}		&
\colhead{N}		&
\colhead{$\Delta$ [Fe/H]}&
\colhead{$\sigma$}	}
\startdata
ZW84  & 15 & $+0.013$ & 0.123 \\
Rut97 & 12 & $-0.014$ & 0.122 \\
CG97  &  6 & $+0.090$ & 0.121 \\
KI03  &  9 & $-0.076$ & 0.110 \\
\enddata
\label{tab:calib}
\end{deluxetable}

Thus, although we can reproduce the metallicities of our calibration
stars to better then $\pm 0.1$ dex, in Table~\ref{tab:bestvalues} we
formally quote our uncertainty as $\pm 0.2$ dex to remind the reader
that this technique is only as good as the metallicity scale of the
calibrating clusters.

\begin{deluxetable}{lc}
\tablewidth{0pt}
\tablecaption{Best Values}
\tabletypesize{\footnotesize}
\tablehead{
\colhead{Cluster}	&
\colhead{[Fe/H]}	}
\startdata
NGC 6256  & $-1.35 \pm 0.20$ \\
NGC 6539  & $-0.79 \pm 0.20$ \\
HP 1      & $-1.30 \pm 0.20$ \\
Liller 1  & $-0.36 \pm 0.20$ \\
Palomar 6 & $-0.52 \pm 0.20$ \\
Terzan 2  & $-0.87 \pm 0.20$ \\
Terzan 4  & $-1.62 \pm 0.20$ \\
\enddata
\label{tab:bestvalues}
\end{deluxetable}

\section{Notes on Individual Clusters} \label{sec:notes}

\subsection{NGC~6256} \label{sec:ngc6256}

\citet{Bica1998} have determined \zzsun\ for a number of metal rich clusters 
based on measurements of the infrared Ca~{\sc ii} triplet in integrated
spectra.  However, due to $\alpha$-element enhancement in metal-rich
clusters, their $Z$ values are expected to be greater than the true
[Fe/H].  For NGC~6256 they derive \zzsun = $-1.01 \pm 0.15$, but also
note that this may only be an upper limit due to contamination of the
integrated spectrum by metal-rich bulge stars.

\citet{Ortolani1999} measured the optical CMD of NGC~6256, and noted 
that it appears to have an extended blue horizontal branch and an RGB
similar to that of NGC~6752 ([Fe/H]$=-1.54$).  Based on these features,
and taking into account the high Bica \etal\ value, they estimate
[Fe/H]$\approx -1.3$.


Using infrared spectra of five giants, we apply the metallicity
solutions based on only the equivalent widths as there is no calibrated
photometry for this cluster.  LL3 gives $-1.34$ with a dispersion of
0.04, and QQ3 gives $-1.35$ with a dispersion of 0.05.  In this case the
error of the mean metallicity will not be set by the dispersion of
observed values ($\sigma/N^{1/2}$), but rather by the accuracy of the
metallicity solution (see Table~\ref{tab:solutions}).  We therefore
choose the QQ3 solution, which has a slightly smaller estimated error,
and assign [Fe/H]$=-1.35 \pm 0.09$ to NGC~6256.

Ortolani's estimate of [Fe/H]$\approx -1.3$ is in excellent agreement
with our value, and the Bica \etal\ measurement is also within range of
our value given their quoted caveats (i.e. that their value of $-1.01$
is an upper limit).  However, the value of $-0.70$ listed in the Harris
catalog appears to be in need of revision downward.

\subsection{NGC~6539} \label{sec:ngc6539}

The only two metallicity determinations for NGC~6539 are those of
\citet{Zinn1984}, who found [Fe/H]$=-0.66 \pm 0.15$ based on integrated
optical spectra, and \citet{Webbink1985}, who estimated [m/H]$=-1.05 \pm
0.28$ from dereddened subgiant colors.


Our new measurement is based on spectra of four stars in NGC~6539.  As
there is no calibrated infrared photometry, we measure metallicities
using only the equivalent widths.  The LL3 solution yields $-0.80$ with
a dispersion of 0.14, and the QQ3 solution gives $-0.79$ with a
dispersion of 0.14.  These star-to-star dispersions are very small, and
the error of the mean metallicity is therefore set by the estimated
accuracy of the technique.  Taking the QQ3 value, we assign [Fe/H]$=-0.79
\pm 0.09$ to NGC~6539.

Both the Zinn \& West and Webbink measurements agree with our value, as
well as the Harris catalog value of $-0.66$.

\subsection{HP~1} \label{sec:hp1}

\citet{Webbink1985} first estimated the metallicity of HP~1 to be 
[m/H]$=-1.68 \pm 0.28$ based on dereddened subgiant colors.  Soon
thereafter \citet{Armandroff1988} found the metallicity to be more than
10 times higher, [Fe/H]$=-0.56 \pm 0.12$, from the strength of the
infrared Ca~{\sc ii} triplet in an integrated spectrum of HP~1.
\citet{Minniti1995a} measured an even higher value of [Fe/H]$=-0.30 \pm
0.20$ based on the strength of several Mg and Fe features in six bright
cluster giants (indicated in Figure \ref{fig:hp1}) chosen on the basis
of their position in the IR CMD.  Unfortunately, there is no overlap
between Minniti's sample and our own; however, it is worthwhile to note
that based on his radial velocities, stars c and d are non-members,
while a and b have velocities similar to that of the central grouping of
stars (``cen'').

\citet{Ortolani1997a} claim that these very high metallicities are due 
to contamination by metal-rich bulge giants.  Using $VI$ CMDs they find
that HP~1 has an extended blue HB and a CMD that resembles that of
NGC~6752, and they estimate [Fe/H]$\approx -1.5$ for HP~1.

\citet[][see note under NGC~6256]{Bica1998} agree that contamination 
from metal-rich bulge giants is very important in HP~1.  Their
integrated spectrum of the central $10''$ yields a low metallicity of
\zzsun$= -1.09 \pm 0.15$, while the spectrum covering the whole cluster
yields a much higher metallicity of \zzsun$=-0.37$.

Using a CO filter to isolate cluster members, \citet{Davidge2000} found
that the RGB of HP~1 is very similar to that of M~13 in a $K$, $J-K$
CMD, implying [Fe/H]$\sim -1.6$.


Our infrared spectroscopic analysis uses six stars measured in HP~1.  As
mentioned earlier, we have divided the HP~1 spectra into ``weak'' and
``strong'' groups based on the strengths of their absorption features
(see Figure \ref{fig:hp1spec}).  Identification of these stars in the
Ortolani \etal\ CMD shows that stars 57, 191, and A are very red, with
$V-I \gtrsim 4$, and lie on the curved RGB of the metal-rich bulge.
Stars 46, 435, and B are not as red, with $V-I \lesssim 3$, and appear
to lie on the more vertical RGB of HP~1.  Therefore, we conclude that
the two ``weak'' stars, 46 and 435, are the true HP~1 members, while the
other ``strong'' stars are surrounding bulge stars.

Based on the weak group, consisting of only stars 46 and 435, we
estimate the cluster metallicity using only the measured equivalent
widths: $-1.25$ with a dispersion of 0.13 from the LL3 solution, and
$-1.30$ with a dispersion of 0.11 from the QQ3 solution.  The
dispersions are so small that the error of the mean is determined by the
accuracy of the technique, and we assign a metallicity of [Fe/H]$=-1.30
\pm 0.09$ for HP~1.

It seems clear that HP~1 has a low metallicity, and higher estimates
were due to contamination by metal-rich bulge stars.  The measurements
of Ortolani \etal ($\approx -1.5$), Bica \etal\ ($-1.09 \pm 0.15$), and
Davidge ($\sim -1.6$), as well as the Harris catalog value of $-1.55$,
are all in reasonable agreement with our new measurement.

\subsection{Liller~1} \label{sec:lil1}

\citet{Malkan1982} first estimated the metallicity of Liller~1 to be
just below solar, [Fe/H]$=-0.21$, based on integrated near-IR photometry
\citep[also listed in][]{Zinn1984,Zinn1985}.  \citet{Armandroff1988}
later raised the value to above solar, [Fe/H]$=+0.20 \pm 0.12$, from
their measurement of the integrated Ca~{\sc ii} triplet.

Studies based on the CMD morphology seemed to confirm the near-solar, or
even supersolar metallicity of Liller~1.  The optical RGB measured by
\citet{Ortolani1996} shows strong curvature and indicates that Liller~1
is considerably more metal-rich than NGC~6528, i.e. [Fe/H]$>-0.17$.  The
slope of the giant branch in the infrared also gives a high metallicity,
[Fe/H]$=+0.25 \pm 0.30$ \citep{Frogel1995}, although this is an
extrapolation of an empirical relation that has only been calibrated
for [Fe/H]$<-0.3$.

The first suggestion of a significantly sub-solar metallicity came from
medium-resolution $H$ and $K$ band spectra of the central $4.4'' \times
6.6''$ of Liller~1 obtained by \citet{Origlia1997}.  They calculated
[Fe/H] for the integrated light based on the strength of the CO(6,3)
\footnote{They point out that the CO bands in the $H$ band are better
indicators of [Fe/H] because they are not as heavily saturated as those
in the $K$ band, and are thus more sensitive to abundance and less
sensitive to the microturbulent velocity.}
absorption feature at 1.62 $\mu$m and found [Fe/H] = $-0.29 \pm 0.3$.
Although soon thereafter \citet[][see note under NGC~6256]{Bica1998}
derived \zzsun$=+0.08 \pm 0.15$ for this cluster based on the integrated
Ca~{\sc ii} triplet.

Most recently, \citet{Davidge2000} has estimated that the metallicity of
Liller~1 is comparable to, or slightly less than that of NGC~6528
([Fe/H]$\lesssim -0.17$) based on filter determined CO band strengths
and infrared colors, and \citet{Origlia2002} have used high-resolution
infrared $H$ band spectra of two bright giants to measure an abundance
of [Fe/H]$=-0.3 \pm 0.2$.


We have obtained spectra of eight stars in Liller~1.  Using only the
equivalent widths, we find a metallicity of $-0.38$ with a dispersion of
0.29 with LL3 and $-0.31$ with a dispersion of 0.28 with QQ3.  However,
including the stellar luminosities and colors from \citet{Frogel1995},
assuming$(m-M)_0=14.91$ and $E(B-V)=3.06$ \citep{Harris1996}, we obtain
$-0.36$ with a dispersion of 0.25 with LLA9 and $-0.30$ with a
dispersion of 0.29 with QQA9.  These values are all consistent with one
another, and we choose the LLA9 solution, which has the smallest
dispersion, giving [Fe/H]$=-0.36 \pm 0.09$ as our best metallicity
measurement for Liller~1.

The recent medium- and high-resolution spectroscopic measurements of
individual giants by Origlia \etal\ are in perfect agreement with our
value, although the estimates based on the CMD morphology or integrated
spectra seem to give values which are too high.  Harris's value of
$+0.22$ also needs revision downwards.

\subsection{Palomar~6} \label{sec:pal6}

The first estimate of the metallicity of Palomar~6 came from the
integrated infrared photometry of \citet{Malkan1982}, who found
[Fe/H]$=-0.74$.  Over a decade later, \citet{Ortolani1995} published the
first optical CMD and estimated that its metallicity is comparable to
NGC~6356, i.e. [Fe/H]$\sim-0.4$, based on the GB morphology.

The highest metallicity estimates come from \citet{Minniti1995a} and
\citet{Bica1998}.  Minniti measured [Fe/H]=$+0.20 \pm 0.30$ based on
optical spectroscopy of four giants selected via IR photometry and
marked with letters b,c,d,f in Figure \ref{fig:pal6}.  Bica \etal\
found \zzsun$=-0.09 \pm 0.15$ based on integrated spectra of the
infrared Ca~{\sc ii} triplet.

Most recently, \citet{Lee2002} measured a very low [Fe/H]$=-1.19 \pm
0.18$ based on the slope of the infrared GB, and $-1.08 \pm 0.06$ from
high-resolution IR spectra of three giants.  They also measured radial
velocities from high-resolution IR echelle spectra of seven giants,
marked with letters a-f in Figure \ref{fig:pal6}.  Their radial
velocities indicate that star 2(g) is a member, while stars 1(f) and
3(b) are nonmembers.  We thus exclude stars 1 and 3 from our analysis of
the cluster metallicity.



As no calibrated infrared photometry is available for Pal~6, our
metallicity determinations are based on the solutions that use only the
equivalent width measurements.  With the five stars measured in
Palomar~6, we find $-0.47$ with a dispersion of 0.44 using LL3 and
$-0.52$ with a dispersion of 0.25 using QQ3.  We choose the quadratic
solution for our best value, as it has the smallest dispersion among the
individual stars, assigning a metallicity of [Fe/H]$=-0.52 \pm 0.11$ to
Palomar~6.

This cluster has an exceptionally large range of metallicity
measurements, ranging from $-1.2$ to $+0.2$, where the Harris catalog
has recently adopted a value of $-1.09$, primarily from \citet{Lee2002}.
Our measurement of $-0.52 \pm 0.11$ falls in the middle of this range,
but is much higher than that measured by \citet{Lee2002}.  One
possibility is that more of our stars are non-members, belonging instead
to the metal-rich bulge population.  If we make the assumption that
stars 5 and 11, which have the strongest absorption features (see
Table~\ref{tab:indices}), are also non-members, keeping only stars 2, 4,
and 14, the resulting metallicity would be $-0.7 \pm 0.2$ (LL3), nearly
consistent with that obtained by \citet{Lee2002}.

\subsection{Terzan~2} \label{sec:ter2}

\citet{Malkan1982} first estimated the metallicity of Terzan~2 to be
[Fe/H]$=-0.47$ from integrated infrared photometry \citep[see
also][]{Zinn1984, Zinn1985}.  The ensuing two studies both found
Terzan~2 to have a slightly higher metallicity of $-0.25$.
\citet{Armandroff1988} measured [Fe/H]$=-0.25 \pm 0.12$ from integrated
spectroscopy of the Ca~{\sc ii} triplet.  \citet{Kuchinski1995} measured
[Fe/H]$=-0.25 \pm 0.25$ based on the slope of the RGB in their infrared
CMD.

\citet{Ortolani1997b} estimated a slightly lower metallicity from their
optical CMD.  Measuring the difference in $V$ magnitude between the HB
and the brightest part of the RGB, they concluded that the metallicity is
between that of 47~Tuc and NGC~6356, or [Fe/H]$\sim -0.55$.

The most recent measurement is that of \citet{Bica1998}, who find
\zzsun$=-0.26 \pm 0.15$ from integrated optical spectra (see note under
NGC~6256).



From our infrared spectra, using only the equivalent widths measured in
seven stars, we find a metallicity of $-0.75$ with a dispersion of 0.38
(LL3) and $-0.83$ with a dispersion of 0.27 (QQ3).  Including the
luminosities and colors from \citet{Kuchinski1995}, assuming
$(m-M)_0=14.70$ and $E(B-V)=2.35$ \citep{Harris1996}, we find $-0.77$
with a dispersion of 0.40 (LLA9) and $-0.71$ with a dispersion of 0.45
(QQA9).  




Looking at the distribution of individual stellar metallicities (last
column of Table~\ref{tab:indices}), star 3 is a $\gtrsim 2 \sigma$
outlier, regardless of which solution is employed.  If we disregard star
3, assuming that it is a non-member, the dispersion of the metallicities
is reduced significantly, and the error of each solution becomes the
limiting factor determining the uncertainty in the metallicity of the
cluster.  We therefore choose the QQA9 solution as the ``best'' and
assign a metallicity of [Fe/H]$=-0.87 \pm 0.06$ to Terzan~2.

Our measurement is the lowest to date, significantly lower than the
Harris catalog value of $-0.40$.  Only the work of Ortolani \etal\ is
consistent with our determination.

\subsection{Terzan~4} \label{sec:ter4}

Because of its crowded, metal-rich environment of the Galactic bulge,
integrated studies of Terzan~4 tend to yield artificially high
metallicities.  Based on integrated infrared photometry
\citet{Malkan1982} measured [Fe/H]$=-0.21$.  Using integrated
spectroscopy of the infrared Ca~{\sc ii} triplet \citet{Armandroff1988}
measured [Fe/H]$=-0.94 \pm 0.12$, and from the integrated spectra of
Ter~4 \citet{Bica1998} found \zzsun$=-0.61 \pm 0.15$.

However, studying the optical CMD of Terzan~4, \citet{Ortolani1997}
point out its similarity to M30, in particular the presence of a blue
horizontal branch, which implies that Terzan~4 could be as metal-poor as
[Fe/H]$\sim -2$.


Our spectroscopic analysis, using only the equivalent widths of the
seven measured stars, yields metallicities of $-1.42$ with a dispersion
of 0.15 (LL3) and $-1.53$ with a dispersion of 0.28 (QQ3).  Including
colors and magnitudes from Frogel \& Sarajedini (1998, private
communication), assuming a distance modulus of $(m-M)_0= 14.795$ and an
extinction of $E(B-V)=2.35$ \citep{Harris1996}, we find $-1.62$ with a
dispersion of 0.21 (LLA9) and $-1.60$ with a dispersion of 0.23 (QQA9).
The error of the mean of the LL3 solution is limited by the accuracy of
the solution (0.11 dex), and therefore the LLA9 value gives the smallest
error, and we assign [Fe/H]$=-1.62\pm 0.08$ to Terzan~4.

The Harris catalog value of $-1.60$ is in perfect agreement with our new
measurement, and the Ortolani \etal\ estimate of $\gtrsim -2$ is also
consistent.


\section{$\omega$~Centauri} \label{sec:wcen}

It has been long known that $\omega$~Centauri is unique among Galactic
globular clusters in that it exhibits a substantial spread in stellar
abundances.  This was first seen in the breadth of the giant branch in
early optical \citep{Woolley1966, Dickens1967, Cannon1973,
LloydEvans1977} and infrared \citep{Persson1980} color-magnitude
diagrams.  Later spectroscopic studies of individual giants
\citep{Cohen1981, Mallia1981, Gratton1982, Paltoglou1989, Brown1993,
Norris1995, Suntzeff1996} showed the range to be approximately one dex.
The most recent photometric and spectroscopic work shows evidence for
three subpopulations in $\omega$~Cen. The largest is a metal-poor group
with a metallicity peak at [Fe/H] $\sim -1.6$, a moderately populated
intermediate-metallicity group at [Fe/H]$\sim -1.2$, and a small number
of metal-rich stars with [Fe/H]$\leq -0.5$, which make up about $\sim$
5\% of the population \citep{Norris1996, Pancino2000, Origlia2003}.

\begin{figure}[htb]
\epsscale{1.2}
\plotone{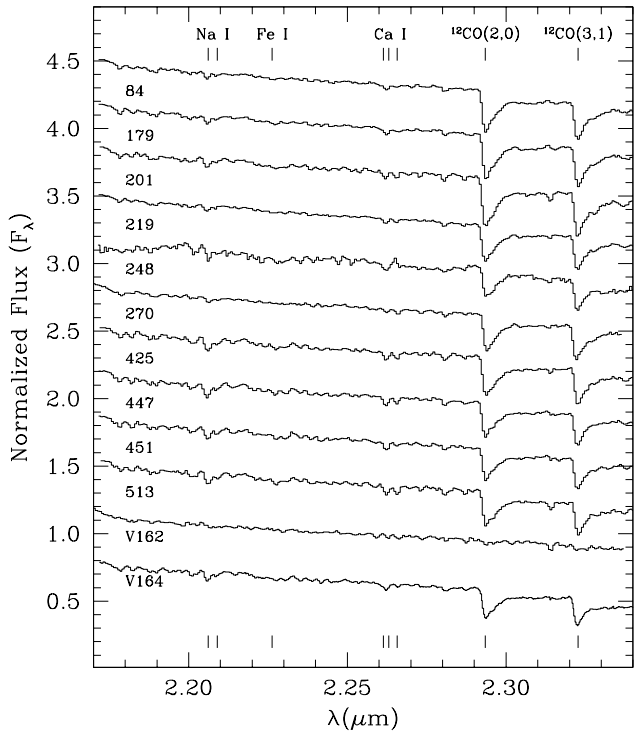} 
\figcaption{
Spectra of the 12 $\omega$~Centauri giants.
V164 is the average of two spectra taken on different nights.
\label{fig:wcenspec}}
\end{figure}

We have obtained infrared spectra of 12 $\omega$~Centauri giants,
selected from the photometry of \citet{Persson1980}.
Figure~\ref{fig:wcenspec} shows the individual spectra, indicating the
wavelengths of some of the strongest absorption features.  A mere visual
inspection of these spectra reveals that most are quite metal-rich,
while V162 is quite metal-poor.

Table~\ref{tab:wcen} lists the infrared photometry from
\citet{Persson1980} assuming $(m-M)_0=13.62$ and $E(B-V)=0.12$
\citep{Harris1996}, measured equivalent widths, and derived
metallicities using each of the four equations from Paper~I.  We note
that several of our stars appear to fall into the ``anomalously
metal-rich'' population, which only accounts for $\sim 5$\% of cluster
\citep{Origlia2003}, while V162 is the only star measured from the
predominant metal-poor population.

\begin{deluxetable*}{ccccccccccccc}
\tablewidth{0pt}
\tablecaption{Observations and Analysis of $\omega$~Centauri Giants}
\tabletypesize{\footnotesize}
\tablehead{
\colhead{}									&
\multicolumn{2}{c}{\underline{ \hspace{2mm} Photometry\tnm{b} \hspace{1mm} }}	&
\multicolumn{3}{c}{\underline{ \hspace{9mm} EW \hspace{8mm} }} 			&
\multicolumn{4}{c}{\underline{ \hspace{14mm} [Fe/H]$_{IR}$ \hspace{13mm} }} 	&
\multicolumn{3}{c}{\underline{ \hspace{7mm} [Fe/H]$_{optical}$ \hspace{6mm} }}	\\
\colhead{Star\tnm{a}}	&
\colhead{$M_{K0}$}	&
\colhead{$(J-K)_0$}	&
\colhead{Na}		&
\colhead{Ca}		&
\colhead{CO}		&
\colhead{LL3}		&
\colhead{LLA9}		&
\colhead{QQ3}		&
\colhead{QQA9}		&
\colhead{ND95\tnm{c}}	&
\colhead{SK96\tnm{d}}	&
\colhead{other}		}
\startdata
  84 & -5.78 & 0.85 &  1.14 &  1.27 & 17.56 & -0.91 & -0.97 & -0.98 & -0.96 &  -1.36 &  -1.16 &  -1.34\tnm{e} \\
 179 & -5.66 & 0.92 &  1.50 &  1.64 & 19.12 & -0.78 & -0.78 & -0.86 & -0.74 &  -1.10 &  -0.80 & \ldots \\
 201 & -5.71 & 0.93 &  1.25 &  1.94 & 21.80 & -0.74 & -0.75 & -0.90 & -0.69 &  -0.85 & \ldots & \ldots \\
 219 & -5.20 & 0.85 &  0.83 &  1.21 & 15.95 & -1.01 & -0.99 & -1.09 & -0.94 &  -1.25 &  -1.01 &  -1.39\tnm{f} \\
 248 & -5.42 & 0.94 &  0.89 &  0.93 &  9.39 & -1.19 & -1.12 & -1.19 & -1.08 &  -0.78 &  -1.03 & \ldots \\
 270 & -4.81 & 0.79 &  0.81 &  1.33 & 15.35 & -1.03 & -0.99 & -1.10 & -0.91 &  -1.22 &  -1.13 &  -1.58\tnm{g} \\
 425 & -5.24 & 0.98 &  2.21 &  2.45 & 16.89 & -0.67 & -0.56 & -0.67 & -0.49 & \ldots & \ldots & \ldots \\ 
 447 & -5.13 & 0.99 &  2.53 &  2.12 & 14.56 & -0.69 & -0.54 & -0.67 & -0.47 & \ldots &  -0.25 & \ldots \\ 
 451 & -4.50 & 0.87 &  1.58 &  2.01 & 12.99 & -0.91 & -0.77 & -0.89 & -0.57 & \ldots & \ldots & \ldots \\ 
 513 & -5.08 & 1.01 &  2.02 &  2.67 & 14.51 & -0.75 & -0.61 & -0.71 & -0.52 & \ldots & \ldots &  -0.90\tnm{h}\\ 
V162 & -5.54 & 0.75 &  0.40 &  0.81 &  0.55 & -1.53 & -1.61 & -1.65 & -1.69 & \ldots & \ldots & \ldots \\ 
V164\tnm{i}& -5.44 & 0.93 & 1.61 & 1.58 & 12.80 & -0.93 & -0.88 & -0.91 & -0.85 & \ldots & \ldots & \ldots \\
\enddata
\tnt{a}{Star IDs are from the ROA catalog \citep{Woolley1966}}
\tnt{b}{Photometry is from \citet{Persson1980}}
\tnt{c}{\citet{Norris1995} from high resolution optical spectra.}
\tnt{d}{\citet{Suntzeff1996}, based on measurements of the Ca {\sc ii} triplet lines (ZW calibration).}
\tnt{e}{\citet{Brown1993} from high resolution optical spectra.}
\tnt{f}{\citet{Gratton1982} from high resolution optical spectra.}
\tnt{g}{\zzsun from \citet{Francois1988} from high resolution optical spectra.}
\tnt{h}{\citet{Origlia2003} from medium resolution infrared spectra.}
\tnt{i}{V164 is the average of two spectra taken on different nights.}
\label{tab:wcen}
\end{deluxetable*}

Previous metallicity determinations of the stars measured here are also
listed in the last three columns of Table~\ref{tab:wcen}.  The
\citet{Norris1995} values were determined from high resolution optical
echelle spectra.  The \citet{Suntzeff1996} values are based on
measurements of the red Ca~{\sc ii} triplet calibrated from the line
strengths measured in Galactic globulars (ZW calibration; an alternate
calibration is based on 24 stars in common with \citet{Norris1995}).
The ``other'' values are explained in the table notes.

On average, there is reasonable agreement between our values and those
previously published.  Taking the average of the four IR solutions
listed in Table~\ref{tab:wcen}, the mean difference for the six stars in
common with \citet{Norris1995} is $-0.15 \pm 0.27$ dex.  Comparing with
\citet{Suntzeff1996}, the mean difference is either $0.02 \pm 0.19$ or
$-0.16 \pm 0.11$, depending on whether we use their ZW or NDC
calibration, respectively.

However, when comparing with the four ``other'' metallicity
determinations listed in the last column of Table~\ref{tab:wcen}, most
of which are from high-resolution spectroscopy, we seem to measure
systematically higher values; the mean difference is $0.40 \pm 0.13$
dex.  At this time it is unclear what could be causing this offset,
although it is most likely {\em not} due to $\alpha$-enhancement, since
the stars in $\omega$ Cen have normal $\alpha$-enhancement compared with
other globular clusters \citep[e.g.][]{Pancino2002}.


Looking at the observed relations of the measured equivalent widths in
$\omega$~Centauri, for example EW(CO) versus EW(Na), EW(CO) versus
EW(Na), and EW(Ca) versus EW(Na), we find that they are completely
consistent with what was seen in the calibrating clusters of Paper~I
(see for example Figure~6 in Paper~I).  This indicates that, at least to
within our measurement accuracy, the abundance ratios seen in these
$\omega$~Cen giants are not vastly different from what we measured for
our calibrating Galactic clusters.  This is consistent with what has
been measured for various elements using high-resolution spectroscopy
\citep[e.g.][]{Pancino2002, Origlia2003}.

\subsection{CO} \label{sec:co}

In Paper~I we observed a tight relationship between the spectroscopic CO
equivalent width and the photometric CO index.  Taking the measurements
of the photometric CO index from \citet{Persson1980}, we find the same
relation also holds true in $\omega$~Centauri.  

\begin{figure}[htb]
\epsscale{1.2}
\plotone{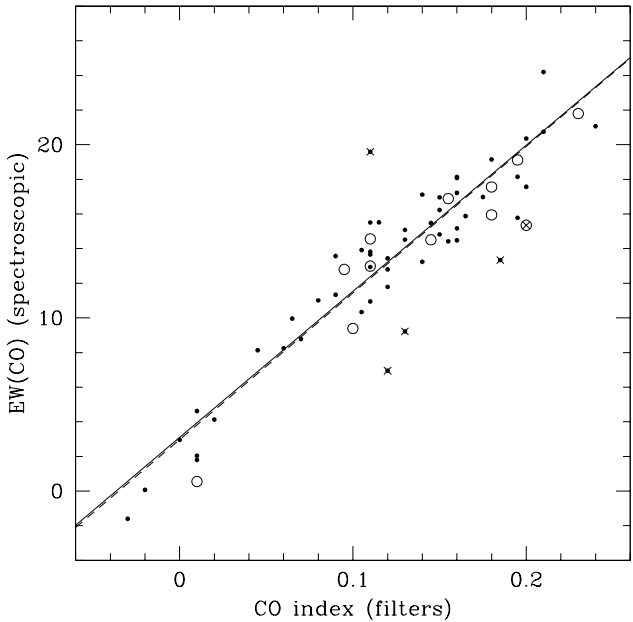} 
\figcaption{
Relationship between the spectroscopically determined EW(CO) and the
photometric CO index for the stars measured in Paper~I (dots) and
$\omega$~Cen (open circles).  The solid line shows the best-fit
least-squares relation for the stars measured in Paper~I only, and the
dashed line is the best-fit relation including the stars measured in
$\omega$~Cen, where in both cases stars with overplotted $\times$'s are
$>2.5 \sigma$ outliers, which are not included in the fits (three of
which are known large-amplitude variables).
\label{fig:wcenco}}
\end{figure}

Figure~\ref{fig:wcenco} shows the spectroscopic EW(CO) as a function of
the photometric CO index for all the stars measured in Paper~I (dots)
and the stars measured in $\omega$~Cen (open circles).  Both sets of
data clearly follow the same relationship and are well modeled by a
simple linear least-squares fit.  The solid line is the fit to only the
data from Paper~I, while the dashed line is to both data sets, except
for the stars with overplotted $\times$'s, which are greater than $2.5
\sigma$ outliers and are not included in the fits.  The fit that
includes both data sets has slightly smaller errors than what was
derived in Paper~I, and is given in Equation~\ref{eqn:co}.

\begin{equation}
  EW(CO) = 2.97 (\pm 0.47) + 84.7 (\pm 3.4) \times CO(index)
\label{eqn:co}
\end{equation}

\section{Discussion \& Conclusions} \label{sec:conclusions}

In this paper we present new metallicities for seven heavily reddened
bulge globular clusters using medium-resolution infrared spectroscopy.
Table~\ref{tab:bestvalues} summarizes these results.  These clusters are
some of the most difficult to study in the Galaxy, and comparing our new
measurements with previous values, we find several systematic trends as
well as a few significant discrepancies.

\begin{figure}[htb]
\epsscale{1.2}
\plotone{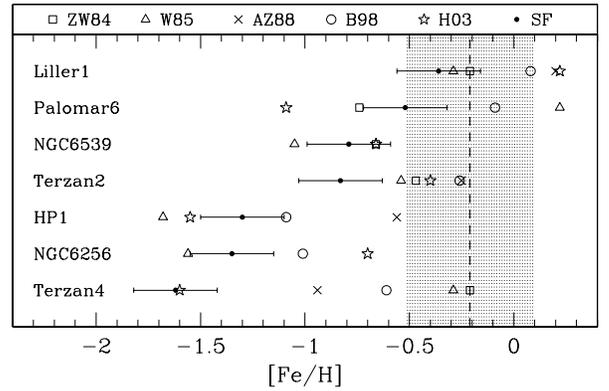} 
\figcaption{
Comparison between previous metallicity determinations and our final
[Fe/H] values determined in Section \ref{sec:notes}.  The dots with
error bars show our measurements (SF) and their absolute uncertainties.
The previous measurements have been taken from Table~\ref{tab:clusters},
although for simplicity we have omitted their error bars.  The vertical
dashed line and shaded region illustrates the mean metallicity and the
measured dispersion of the Galactic bulge \citep[$-0.21 \pm
0.30$;][]{Ramirez2000}.
\label{fig:fehcomparison}}
\end{figure}

Figure~\ref{fig:fehcomparison} illustrates these differences and trends.
Here we plot our new metallicity determinations from
Table~\ref{tab:bestvalues} with those from previous works listed in
Table~\ref{tab:clusters}.  The dots represent our measurements, where
the uncertainties are taken to be $\pm 0.2$ dex, as discussed in
Section~\ref{sec:fehscale}.

The \citep{Bica1998} measurements are based on the strength of the red
Ca~{\sc ii} triplet in the integrated cluster light.  However, since
Galactic globulars are $\alpha$-enhanced \citep[see][for a review of
$\alpha$-enhancement in globular clusters]{Carney1996}, the \zzsun
values of \citet{Bica1998}, which measure an overall metallicity, will
be slightly higher than the iron abundance, [Fe/H].  Indeed, comparing
with our final abundances, we find that in the mean, the
\citet{Bica1998} values are systematically higher by $0.50 \pm 0.28$ dex
(see last row of Table~\ref{tab:clusters}).

The \citet[][AZ88]{Armandroff1988} measurements, also based on the
strength of the red Ca~{\sc ii} triplet, also give a large deviation
from our final values: $0.64 \pm 0.08$ dex.  There are several possible
explanations for this large discrepancy.  AZ88 calibrated their summed
Ca~{\sc ii} equivalent width ($\Sigma W$) using a linear fit between
$\Sigma W$ and [Fe/H], as measured by \citet{Zinn1984} for seven
clusters.  However, it is now known that the strength of the Ca~{\sc ii}
lines shows a non-linear relationship with [Fe/H]$_{ZW84}$, especially
for [Fe/H]$_{ZW84} \gtrsim -0.7$ \citep[][see also Section
\ref{sec:fehscale}]{Rutledge1997, Kraft2003}.  This poor calibration
could certainly lead to overestimates of the metallicity; however, the
most likely explanation is simply contamination from surrounding
metal-rich bulge stars.

The Harris catalog \citep{Harris1996} is an average of globular cluster
parameters taken from the literature.  It is maintained on the web at
http://physun.physics.mcmaster.ca/Globular.html.  While originally on
the \citet{Zinn1984} scale, which appears slightly nonlinear and tends
to overestimate the metallicities of the most metal-rich clusters, more
recent versions of the catalog have incorporated many values from
high-resolution spectroscopy and are thus closer to the
\citet{Carretta1997} scale (see Section \ref{sec:fehscale} for a
discussion of the various metallicity scales, and how our technique
compares).

In general, the comparison of our values with those in the 2003 version
of the Harris catalog is very close; the mean difference between the two
sets of measurements is $+0.14 \pm 0.45$ dex (last line of
Table~\ref{tab:clusters}), where the Harris values are slightly higher
than ours.  The explanation behind the higher metallicities found in
previous studies seems to be primarily contamination by metal-rich bulge
stars (see Figure~\ref{fig:fehcomparison}).  The combination of very
crowded fields, low-concentration clusters, and similar stellar
luminosities makes it very difficult to distinguish between cluster
members and superposed bulge stars.

The largest discrepancies between our new values and those in the Harris
catalog are those of NGC~6256 and Liller~1, which are listed with
metallicities $0.65$ and $0.59$ dex higher than we measure, respectively.
Our lower values for both clusters are in good agreement with other
recent studies.  The Harris catalog also lists Palomar~6 with a
metallicity of $-1.09$, taken from the IR spectra of \citet{Lee2002},
which is $0.57$ dex {\em lower} than we measure.  See Section
\ref{sec:pal6} for a discussion of this cluster.

As a result of this study we have reduced by three the number of
Galactic globular clusters that have [Fe/H]$\gtrsim -0.7$ according to
the Harris catalog.  The metallicity of NGC~6256 was lowered to $-1.35$
and NGC~6539 was decreased down to $-0.79$.  Terzan~2, which lies within
10\degr\ of the Galactic center, was reduced to [Fe/H]=$-0.87$.

We have also observed 12 giants in the globular cluster
$\omega$~Centauri, all except one of which appear to lie in the
intermediate- or metal-rich regions of $\omega$~Cen's tri-modal
metallicity distribution.  We compare our results for the eight stars
that have previous metallicity determinations and find general
agreement, although some of the most recent optical spectroscopy finds
slightly lower metallicities than we predict.


\acknowledgements 

Support for this work was provided by a Princeton-Catolica Prize Fellowship, 
and Proyecto FONDECYT Regular No. 1030976, both awarded to AWS.
This paper was based on spectroscopic observations obtained at the Cerro
Tololo Interamerican Observatory.
Thanks to Ata Sarajedini for collaborating on obtaining photometry of Terzan~4.
Thanks to Karrie Gilbert and Nathan Hostler for their assistance
obtaining some of the photometric observations at the MDM observatory,
which were used for target selection.
Thanks to Dante Minniti for providing his unpublished finder charts for
Pal~6 and HP~1.
JAF thanks Sean Solomon for his hospitality at DTM.
We would also like to thank the {\em two} anonymous referees who
provided very useful comments which greatly improved this paper.
The ESO/ST-ECF Science Archive Facility was used to retrieve
observations made with the European Southern Observatory telescopes in
order to construct the $I$ band finder charts for NGC~6256, Terzan~4,
and HP~1.

\end{document}